\documentclass[aps,twocolumn,preprintnumbers,prd,superscriptaddress,10pt]{revtex4-2}

\usepackage{amssymb,amsmath,amsthm,amsfonts}
\usepackage[usenames]{color}
\definecolor{darkred}{rgb}{0.5,0,0}

\usepackage{bm}
\usepackage{dcolumn}
\usepackage[utf8]{inputenc}
\usepackage{latexsym}
\usepackage{rotating}
\usepackage{longtable}

\usepackage{enumerate}
\usepackage{tensor,multirow}
\usepackage{url}
\usepackage[linktocpage]{hyperref}
\usepackage{float}
\usepackage{graphicx}

\begin{document}

\title{Kicks in charged black hole binaries}

\author{Raimon~Luna}

\affiliation{Departament d'Astronomia i Astrof\'{\i}sica, Universitat de Val\`encia,
Dr.\ Moliner 50, 46100, Burjassot (Val\`encia), Spain}
\affiliation{Departament de F{\'\i}sica Qu\`antica i Astrof\'{\i}sica, Institut de Ci\`encies del Cosmos,
Universitat de Barcelona, Mart\'{\i} i Franqu\`es 1, E-08028 Barcelona, Spain}
\affiliation{Centro de Astrof\'\i sica e Gravita\c c\~ao -- CENTRA,
  Departamento de F\'\i sica, Instituto Superior T\'ecnico -- IST,
  Universidade de Lisboa -- UL, Avenida Rovisco Pais 1, 1049-001 Lisboa, Portugal}

\author{Gabriele~Bozzola}
\affiliation{Department of Astronomy, University of Arizona, Tucson, AZ, USA}

\author{Vitor~Cardoso}
\affiliation{Centro de Astrof\'\i sica e Gravita\c c\~ao -- CENTRA,
  Departamento de F\'\i sica, Instituto Superior T\'ecnico -- IST,
  Universidade de Lisboa -- UL, Avenida Rovisco Pais 1, 1049-001 Lisboa, Portugal}
\affiliation{Niels Bohr International Academy, Niels Bohr Institute, Blegdamsvej 17, 2100 Copenhagen, Denmark}

\author{Vasileios~Paschalidis}
\affiliation{Department of Astronomy, University of Arizona, Tucson, AZ, USA}
\affiliation{Department of Physics, University of Arizona, Tucson, AZ, USA}

\author{Miguel~Zilh\~ao}
  \affiliation{Departamento  de  Matem\'{a}tica  da  Universidade  de  Aveiro and 
  Centre for Research and Development in Mathematics and Applications (CIDMA),
  Campus de Santiago, 3810-183 Aveiro, Portugal}
\affiliation{Centro de Astrof\'\i sica e Gravita\c c\~ao -- CENTRA,
  Departamento de F\'\i sica, Instituto Superior T\'ecnico -- IST,
  Universidade de Lisboa -- UL, Avenida Rovisco Pais 1, 1049-001 Lisboa, Portugal}

\begin{abstract}
We compute the emission of linear momentum (kicks) by both gravitational and electromagnetic radiation in fully general-relativistic numerical evolutions of quasi-circular charged black hole binaries. We derive analytical expressions for slowly moving bodies and  explore numerically a variety of mass ratios and charge-to-mass ratios. We find that for the equal mass case our analytical expression is in excellent agreement with the observed values and, contrarily to what happens in the vacuum case, we find that in presence of electromagnetic fields there is emission of momentum by gravitational waves. We also find that the strong gravitational kicks of binaries with unequal masses affect the electromagnetic kicks, causing them to strongly deviate from Keplerian predictions. For the values of charge-to-mass ratio considered in this work, we observe that magnitudes of the electromagnetic kicks are always smaller than the gravitational ones.
\end{abstract}

\maketitle

\section{Introduction}
\label{Introduction}
With the advent of gravitational-wave (GW) astronomy~\cite{LIGOScientific:2018mvr}, the last decade has seen remarkable progress in our ability to study strong-field gravity in highly dynamical regimes. We can now monitor the GW-driven coalescence of compact binaries, paving the way to new tests of General Relativity (GR) in the strong-field regime~\cite{Yagi:2016jml,Barack:2018yly,Cardoso:2019rvt}. Some of these studies rely on the simplicity of black holes (BHs) in GR, namely on the mathematical result that asymptotically flat vacuum BH geometries must be parameterized by a small number of quantities (their mass, angular momentum, and charge)~\citep{Bekenstein:1971hc,Robinson:1975bv,Chrusciel:2012jk}.
Nevertheless, the anticipated breakdown of classical GR in BH interiors and our incomplete understanding of the matter content of the Universe motivates the search for smoking-gun evidence for beyond-standard-model physics in GW data.

While the search for new physics is challenging (in part due to our nearly total ignorance of a quantum theory of gravity or the nature of dark matter), we can take guidance from robust aspects of specific modified theories of gravity~\cite{Berti:2015itd,Barack:2018yly}. For instance, a generic feature that arise when extra degrees of freedom are involved is the dipole emission~\cite{Barausse:2016eii}. A specific example that we will use as a prototype for modified gravity theories with such effect is Einstein's theory minimally coupled to a massless vector field, known as Einstein-Maxwell's theory. In contrast to most beyond-standard-model theories, Einstein-Maxwell admits a well-posed initial value problem~\cite{Alcubierre2009}, and hence is amenable to numerical integration.
In this theory, BHs can carry a conserved charge. While this charge can be thought of as electric charge, astrophysical BHs are expected to be electrically neutral to a very good approximation. This is due to Schwinger-type and Hawking radiation mechanisms, and the availability of interstellar plasma, mechanisms that are effective largely because of the huge charge-to-mass ratio of electrons (see, e.g.~\cite{Cardoso:2016olt} and references therein).
However, the mathematical description for Einstein-Maxwell does not only describe electric charge but can be applied to any $U(1)$ field.
If the vector field is not a Maxwell field, it is possible to envision scenarios where the charge-to-mass ratio of fundamental particles is smaller, and where BHs could be charged under such a field. It is also possible to interpret the charge as magnetic in nature (possibly due to primordial magnetic monopoles). Within this framework,
the existence of interaction and another channel of radiation, in addition to the GW channel, changes the dynamics of the compact binary, leading to potentially observable signatures in the GWs.
Indeed, the first constraints on the charge and dipole moment of BH binaries, using GW detection, have been placed~\cite{Barausse:2016eii,Cardoso:2016olt,Bozzola:2020mjx,Carullo:2021oxn} and show that data is compatible with non-negligible amount of charge.
In other words, the Einstein-Maxwell theory is a good proxy for more general theories of gravitation, but also an attractive candidate for dark matter~\cite{Cardoso:2016olt} (for theoretical scenarios where Einstein-Maxwell theory is mathematically applicable, see, e.g.,~\cite{Bozzola:2019aaw,Bozzola:2020mjx,Bozzola:2021elc}).

One potentially important, but unexplored, consequence of having charged objects is that EM radiation will induce a recoil on the final object, in addition to that induced by GW emission. The recoil induced by GWs was studied extensively (see Refs.~\cite{Bonnor1961,PhysRev.128.2471,1973ApJ...183..657B,Fitchett,Fitchett:1984qn, Lousto:2004pr, Herrmann:2006cd, Campanelli:2004zw, Baker:2006vn, Gonzalez:2006md, Campanelli:2007cga, Gonzalez:2007hi, Lousto:2011kp} for a very incomplete list). This gravitational recoil, known as the {\it kick}, has important implications on the structure of galaxies and the formation of supermassive BHs~\cite{Boylan-Kolchin:2004fnd, Madau:2004st, Merritt:2004xa, Libeskind:2005eh, Volonteri:2005pn, Haiman:2004ve} and particularly due to the possibility that the resulting BH might exceed the escape velocity of globular clusters, or even galaxies, thus being ejected from them~\cite{Haehnelt:2005bh, Magain:2005yp, Hoffman:2005be, Merritt:2005nt, Merritt:2002hc}. 

The rate of emission of momentum in GWs is caused by the interaction between the quadrupole and the octupole of the energy distribution \citep{Fitchett}. In gravity theories with dipole emission, this suggests that the leading-order effect giving rise to a kick would be a dipole-quadrupole term,
which can lead to qualitatively new features, potentially surpassing the GW effect. The purpose of this work is to explore this question. Since in GR the kick is independent of the scale of the system, our study applies to both stellar mass and supermassive BHs alike. The detection of kicks in GW data~\cite{Gerosa:2016vip,Varma:2022pld} allows one to place additional constraints on dipole emission; these require a proper modeling of the system, an important motivation for this analysis.

The paper is structured as follows: In Section~\ref{sec:Keplerian} we estimate the kicks analytically assuming Keplerian circular orbits. The numerical setup and code are explained in Section~\ref{sec:evolutions}. Section~\ref{sec:Results} exposes the results of the simulations and compares them with the Keplerian estimate. We conclude in Section~\ref{sec:Discussion}. We use units in which $G=c=1$, where $G$ is Newton's constant and $c$ the speed of light in vacuum. The conventions for the EM fields are those of \cite{Zilhao:2015tya}. We express results in terms of the Arnowitt-Deser-Misner (ADM) mass of the system, $M_{ADM}$. Greek indices range from 0 to 3, and Latin ones range from 1 to 3.

\section{Slow motion, weak field approximation}
\label{sec:Keplerian}

We follow here an approach analogous to what Ref.~\cite{1973ApJ...183..657B} did for GW emission. We use a slow-motion, weak-field approximation, and start by solving for the EM field $A^\mu$ in the Lorenz gauge, which leads to a wave equation with source
\begin{equation}
\Box A^\mu ({\bf x}, t) = 4 \pi J ^\mu.
\end{equation}
Here, $\mathbf{x}$ and $t$ are the space and time coordinates in flat space, and $J^{\mu}$ is the EM current density. Using the Green's function of the d'Alembertian operator we get
\begin{equation}
A^\mu({\bf x}, t) =  \int \frac{J^\mu({\bf x}', t - |{\bf x} - {\bf x}'|)}{|{\bf x} - {\bf x}'|} dV
\end{equation}
with $dV = d^3\mathbf{x}'$. We will drop the primes from now on. A straightforward manipulation yields, up to higher order terms
\begin{equation}
\begin{split}
A^i({\bf x}, t) &= \frac{1}{r} \left( \dot D^i - \frac{n^j \dot M^{ij}}{2}  + \frac{n^j \ddot B^{ij}}{2}  \right),\\
B^{ij} &= \frac{1}{3}\left(Q^{ij} + \delta^{ij} \int J^0 r^2 dV\right),
\end{split}
\end{equation}
where $n^i$ is the outward-pointing unit 3-vector, and we define the electric dipole $D^i$, the magnetic dipole $M^{ij}$ and the traceless electric quadrupole $Q^{ij}$ as
\begin{equation}
\begin{split}
D^i &= \int J^0 x^i dV, \\ M^{ij} &= \int( x^i J^j - x^j J^i) dV, \\
Q^{ij} &= \int J^0 (3 x^i x^j - r^2 \delta^{ij}) dV.
\end{split}
\end{equation}
At sufficiently large distances from the source, the fields behave as a plane wave~\cite{Landau}. In this limit, by neglecting the Coulomb term ($\propto r^{-2}$) in front of the radiative term ($\propto r^{-1}$) the Poynting vector becomes simply
\begin{equation}
{\bf S} = |{\bf \dot A} \times {\bf n}|^2 \, \frac{\bf n}{4\pi} . \label{eq:Poynting}
\end{equation}
After a lengthy manipulation of~\eqref{eq:Poynting}, and integrating over solid angle, we are left with the expression for the rate of emission of linear momentum in the EM radiation
\begin{equation}
\frac{dP_{\rm EM}^i}{dt} =
\frac{1}{15} \ddot D^{j} \dddot Q^{ji} - \ddot D^{j} \ddot M^{ji}.
\end{equation}
We see here that the dominant contribution to the momentum emission is a combination of electric dipole and magnetic dipole, and electric dipole and quadrupole terms. This formula is somewhat similar to its GW analog, where the momentum emission arises from the interaction between the mass quadrupole and a combination of mass octupole and angular momentum moments \cite{1973ApJ...183..657B}.

For the particular case of point particles following Keplerian circular orbits, we set~\cite{Fitchett}
\begin{equation}
\begin{split}
J^0 &= \sum_{n=1}^2 q_n \delta({\bf x} - {\bf x}_n(t)), \\ J^i &= \sum_{n=1}^2 q_n \dot x_n^i\delta({\bf x} - {\bf x}_n(t)),
\end{split}
\end{equation}
with $\delta$ being Dirac's delta.
If the particles have masses $m_1$ and $m_2$, charges $q_1$ and $q_2$, and are at a distance $d$ of each other (distance $d_1$ and $d_2$ from the center of mass), a direct modification of Kepler's third law yields an angular velocity
\begin{equation}
\Omega_c = d^{-3/2}\sqrt{(m_1+m_2)\left(1 - \frac{q_1 q_2}{m_1 m_2}\right)}\,,
\end{equation}
where we also accounted for the EM interaction.
Without loss of generality, we may choose an instant of time where both particles lie along the $x$-axis, with their velocity vectors in the $y$ direction. The only nonzero components of the electric dipole and quadrupole tensor derivatives are then
\begin{equation}
\begin{split}
\ddot D^1 &= - (d_1 q_1 - d_2 q_2) \Omega_c^2, \\ \dddot Q^{12} &= \dddot Q^{21} = - 12  (d_1^2 q_1 + d_2^2 q_2) \Omega_c^3\,,
\end{split}
\end{equation}
and the magnetic dipole tensor is zero. Therefore
\begin{align}
\frac{d{\bf P}_{\rm EM}^i}{dt} &= \frac{4}{5} (d_1 q_1 - d_2 q_2) (d_1^2 q_1 + d_2^2 q_2)\; \Omega_c^5 \; \hat {\bf y} \notag \\
 &=  \left(\frac{q_1}{m_1}-\frac{q_2}{m_2}\right)  \left(\frac{q_1}{m_1^2}+\frac{q_2}{m_2^2}\right) F_0\;  \hat {\bf y},
\end{align}
with
\begin{equation}
F_0 \equiv \frac{4}{5} \frac{m_1^3 m_2^3}{\sqrt{(m_1+m_2)d^9}}{ \left(1-\frac{q_1 q_2}{m_1 m_2}\right)}^{5/2}\,.
\end{equation}
By defining
\begin{equation}
M = m_1 + m_2\,,\quad \rho =\frac{m_1}{m_2}\,,\quad \lambda_i = \frac{q_i}{m_i}\,,
\end{equation}
we obtain
\begin{equation}
\frac{dP_{\rm EM}}{dt} = \frac{4}{5} {\left(\frac{M}{d}\right)}^{9/2} \frac{ {\left(1 - \lambda_1 \lambda_2\right)}^{5/2}\rho^2 }{{(1+\rho)}^5}  \left(\lambda_1 - \lambda_2\right) \left(\lambda_1 + \rho \lambda_2\right)\,.
\label{eqn:thrust}
\end{equation}
Following~\cite{Gonzalez:2006md}, we assume that the total integrated momentum will have the same functional form.
This result predicts a zero kick for the cases $\lambda_{1} = \lambda_{2}$ (when the dipole vanishes) and $\lambda_{1} = -\rho\lambda_{2}$ (when the quadrupole vanishes), in the center of mass frame. Note that the factor $\left(1 - \lambda_1 \lambda_2 \right)^{5/2}$ is never zero in BHs since $\lambda_i$ are strictly less than~1. We will now use full nonlinear simulations to compute the actual recoil imparted by EM radiation, trying to make contact with this flat-space result.

\section{Setup}%
\label{sec:evolutions}
The action for the Einstein-Maxwell theory is
\begin{equation}
S = \int d^4 x \sqrt{-g} \left(\frac{R}{4} - \frac{1}{4} F^{\mu\nu}F_{\mu\nu}\right)
\end{equation}
where $R$ is the Ricci scalar, and the electromagnetic field strength is defined as $F_{\mu\nu} = \nabla_\mu A_\nu - \nabla_\nu A_\mu$. Variation with respect to the metric tensor $g_{\mu\nu}$ and the electromagnetic vector potential $A_\mu$ result, respectively, in 
\begin{align}
& \nabla_\nu F^{\mu\nu} = 0, \\
& R_{\mu\nu} - \frac{1}{2} g_{\mu\nu} R = 2 F_{\mu\rho} F_\nu^{\; \rho} - \frac{1}{2} g_{\mu\nu} F^{\rho\sigma}F_{\rho\sigma}.
\end{align}
To evolve these equations numerically we employ the standard 3+1 decomposition to formulate the equations of motion as a Cauchy problem under the Baumgarte-Shapiro-Shibata-Nakamura (BSSN) scheme~\cite{PhysRevD.52.5428, PhysRevD.59.024007}.
Our numerical approach follows that of previous evolutions of charged BHs~\cite{Zilhao:2012gp,Zilhao:2013nda,Zilhao:2014wqa,Bozzola:2020mjx,Bozzola:2021elc,Bozzola:2022uqu}. Specifically, numerical computations are performed using the \texttt{EinsteinToolkit} (ET) code~\cite{EinsteinToolkit:2021_11}, within the \texttt{Cactus} framework~\cite{Goodale:2002a} with grid functions being computed on a \texttt{Carpet} adaptive-mesh-refinement Cartesian grid~\cite{Schnetter:2003rb}. To prepare initial data and diagnose the spacetime we adopt the codes \texttt{TwoChargedPunctures} and \texttt{QuasiLocalMeasuresEM} developed in~\cite{Bozzola:2019aaw}, which extend
the \texttt{TwoPunctures}~\cite{Ansorg:2004ds} and \texttt{QuasiLocalMeasures}~\cite{Dreyer:2002mx} to the full Einstein-Maxwell theory. These codes have been employed in~\cite{Bozzola:2020mjx, Bozzola:2021elc, Bozzola:2022uqu}, where it was demonstrated how one can perform long-term and stable quasi-circular charged BH numerical evolutions. For a different approach to generating initial data, see~\cite{Mukherjee:2022gla}. Time evolution of the electromagnetic fields is performed with the massless version of \texttt{ProcaEvolve} thorn~\cite{Zilhao:2015tya} within the \texttt{Canuda} library~\cite{witek_helvi_2021_5520862}. The spacetime is evolved with \texttt{Lean}~\cite{Sperhake2007}, also within \texttt{Canuda}. We employed the \emph{continuous} Kreiss-Oliger dissipation prescription introduced in~\cite{Bozzola:2021elc}. \texttt{AHFinderDirect}~\cite{Thornburg:2003sf} is adopted for locating apparent horizons. To save computational resources we impose symmetry with respect to the orbital $z=0$ plane. 

We approximate the initial momenta of the  BHs in a quasi-circular inspiral by the 3.5 Post-Newtonian approximation for neutral BHs, and rescaling it by a factor of $\sqrt{1 - \lambda_1 \lambda_2}$ as explained in \cite{Bozzola:2021elc}.

The emission of energy by EM and GWs can be computed from the Newman-Penrose scalars~\cite{Newman1962, Zilhao:2015tya} $\Psi_{4}$ and $\Phi_{2}$, respectively, as
\begin{align}
\frac{dE_{\rm GW}}{dt} &= \lim_{r \to \infty} \frac{r^2}{16 \pi} \oint \left|\int_{-\infty}^t \Psi_4 \, dt'\right|^2 d\Omega\,,\label{eq:Egw}\\
\frac{dE_{\rm EM}}{dt} &= \lim_{r \to \infty} \frac{r^2}{4 \pi} \oint \left|\Phi_2\right|^2 d\Omega\,,\label{eq:Eem}
\end{align}
where $\oint$ indicates surface integration over a coordinate sphere of radius $r$
and $d \Omega = \sin \theta d \theta d \phi$ is the differential solid angle.

The expressions for linear momentum emission are obtained simply by adding a radially-pointing unit vector $\hat n$ inside the angular integral,
\begin{align}
\frac{dP_{\rm GW}^i}{dt} &= \lim_{r \to \infty} \frac{r^2}{16 \pi} \oint n^i \, \left|\int_{-\infty}^t  \Psi_4 \, dt'\right|^2 d\Omega\,,\label{eq:Pgw}\\
\frac{dP_{\rm EM}^i}{dt} &= \lim_{r \to \infty} \frac{r^2}{4 \pi} \oint n^i \, \left|\Phi_2\right|^2 d\Omega.\label{eq:Pem}
\end{align}
The actual computation is done in terms of the multipole decomposition coefficients of $\Psi_4$ and $\Phi_2$ as explained in Appendix~\ref{app:Multipoles}. We use multipole moments up to order $l = 8$, and we start the time integration at a retarded time of 40 computational units (about $39.2 M_{\rm ADM}$) in order to avoid the ``junk'' radiation (spurious radiation, arising from the way that initial data is specified). As explained bellow, we then convert the linear momentum to recoil velocity as $v = P/M_F$, where $M_F$ is the quasi-isolated mass of the final BH as computed by \texttt{QuasiLocalMeasuresEM}.

Several sanity checks (such as the comparison to neutral results of Ref.~\cite{Gonzalez:2006md}) have been performed on the numerical implementation of the equations, and are summarized in Appendix~\ref{app:Checks}. The computation of the kicks from the multipole moments has been compared to the direct integration of momentum emission over solid angle, showing excellent agreement. The computation of the Newman-Penrose scalars is implemented in the ET \texttt{NPScalars\_Proca} thorn~\cite{Zilhao:2015tya}. Each run has 8 refinement levels (each one halving the previous grid spacing), with outermost grid spacing of $\Delta x = \Delta y = \Delta z = 1.23 M_{ADM}$. This corresponds to having at least 50 points across the horizon in cases with $\rho = 1$, and at least 33 across the diameter of the horizon of the smallest BH for $\rho = 2$.

\begin{table}[thpb]
\caption{Parameters used in the initial data generation, expressed in code units, for the first (top) and second (bottom) series of runs.\label{tab:initial_data}}
\setlength\tabcolsep{8pt}
\begin{tabular}{cccccc}
\hline
 $M_{ADM}$ & $m_1$ & $m_2$ & $\rho = m_1/m_2$ & $\lambda_1$ & $\lambda_2$ \\
\hline
 1.02 &  0.52 &  0.52 &  1.00 &  0.19 & -0.19 \\
 1.02 &  0.52 &  0.52 &  1.00 &  0.19 & -0.10 \\
 1.03 &  0.52 &  0.52 &  1.00 &  0.19 &  0.00 \\
 1.03 &  0.52 &  0.52 &  1.00 &  0.19 &  0.10 \\
 1.02 &  0.52 &  0.52 &  1.00 &  0.19 &  0.19 \\
\hline
\hline
 1.02 &  0.68 &  0.35 &  1.94 &  0.19 & -0.19 \\
 1.02 &  0.68 &  0.35 &  1.94 &  0.19 & -0.09 \\
 1.02 &  0.68 &  0.35 &  1.94 &  0.19 &  0.00 \\
 1.02 &  0.68 &  0.35 &  1.94 &  0.19 &  0.09 \\
 1.02 &  0.68 &  0.35 &  1.94 &  0.19 &  0.19 \\
\hline
\end{tabular}
\end{table}

We performed a sampling of a subset of the parameters describing a binary BH. Namely, we use an initial coordinate distance of $d = 6.86 M_{ADM}$, and we organize the runs by ``series'' of fixed mass ratio $\rho = m_1 / m_2$ and $\lambda_1$ ($\rho \geq 1$). For each series, we vary $\lambda_2$ from $-\lambda_1$ to $\lambda_1$ in intervals of 0.1. The construction of the initial data by \texttt{TwoChargedPunctures} fixing the bare masses introduces some small deviations from the target initial masses and charges of the individual BHs, and also to the global ADM mass of the system. This causes the mass ratios to be slightly different from the exact values of $\rho = 1, 2$, and the charge-to-mass ratios $\lambda$ to vary between runs, as measured by \texttt{QuasiLocalMeasuresEM}. Additionally, the fact that $M_{\rm ADM} \neq 1$ creates a discrepancy between the computational code units and the geometric units in terms of the total mass. For this reason, it is important to keep in mind that for all plots and fitting functions that assume constant quantities in several runs, this assumption is only approximately satisfied. Also, waveforms extracted at a fixed radius $R_{\rm ex}$ in computational units will not correspond exactly to the same physical distance, although the difference is negligible compared to other sources of error. Table~\ref{tab:initial_data} lists the ADM mass, physical masses of each BH, charge-to-mass ratios and binary mass ratios that we explored. The ADM mass is computed at the level of initial data by surface integrals as in \cite{baumgarte_shapiro_2010}, and the physical masses $m_{1,2}$ refer to the values measured by \texttt{QuasiLocalMeasuresEM} as in \cite{Bozzola:2019aaw}.

Fits to data were performed by the nonlinear least-squares (NLLS) Marquardt-Levenberg algorithm implementation in \texttt{Gnuplot}, which also provides an estimate of the asymptotic standard errors in the coefficients.

\section{Results}%
\label{sec:Results}

In Table \ref{tab:kicks} we list the values of the kicks obtained for each of the parameter configurations.

\begin{table}[thpb]
\caption{Values of the kicks for the first (top) and second (bottom) series of runs, all of them having $\lambda_1 = 0.19$.\label{tab:kicks}}
\setlength\tabcolsep{10pt}
\begin{tabular}{c|ccccc}
\hline
 $\rho = 1.00$ & $\lambda_2$ & $v_{EM}$ (km/s) & $v_{GW}$ (km/s) \\
\hline
 & -0.19 &  0.00 $\pm$  0.00 &  0.00 $\pm$  0.00 \\
 & -0.10 &  4.73 $\pm$  0.03 &  2.01 $\pm$  0.09 \\
 &  0.00 &  6.25 $\pm$  0.04 &  2.62 $\pm$  0.12 \\
 &  0.10 &  4.66 $\pm$  0.03 &  1.93 $\pm$  0.09 \\
 &  0.19 &  0.00 $\pm$  0.00 &  0.00 $\pm$  0.00 \\
\hline
\hline
 $\rho = 1.94$ & $\lambda_2$ & $v_{EM}$ (km/s) & $v_{GW}$ (km/s) \\
\hline
 & -0.19 &  6.17 $\pm$  0.04 &  153.36 $\pm$  6.9 \\
 & -0.09 &  7.91 $\pm$  0.05 &  149.54 $\pm$  6.7 \\
 &  0.00 &  7.76 $\pm$  0.05 &  146.23 $\pm$  6.6 \\
 &  0.09 &  5.67 $\pm$  0.03 &  144.74 $\pm$  6.5 \\
 &  0.19 &  1.75 $\pm$  0.01 &  144.42 $\pm$  6.5 \\
\hline
\end{tabular}
\end{table}
Although  Eq.~\eqref{eqn:thrust} assumes flat space and a circular trajectory, it suggests a functional form for the EM kick. We thus expect the rate of emission of EM momentum as a function of $\lambda_2$ to be a curve close to a parabola, with roots at $\lambda_2 = \lambda_1$ (zero dipole) and at $\lambda_2 = -\lambda_1 / \rho$ (zero quadrupole). As in Ref.~\cite{Gonzalez:2006md}, we conjecture that the total integrated momentum (the kick) should follow the same kind of dependence. Therefore, we fit a function of the form
\begin{equation}
v_{\rm EM} = A \left(B - \lambda_2\right) \left(C + \lambda_2 \right) {\left(1 - D \lambda_2\right)}^{5/2}.
\label{eq:fit_form}
\end{equation}
Based on the Keplerian expression, we expect $B = \lambda_1$,  $C = \lambda_1 / \rho$ and $D = \lambda_1$. The constant $A$ will depend on the details of the decay of the orbital separation, and is therefore not possible to extract it at a purely Newtonian level.
\begin{figure}[thpb]
\begin{center}
\includegraphics[width=0.43\textwidth]{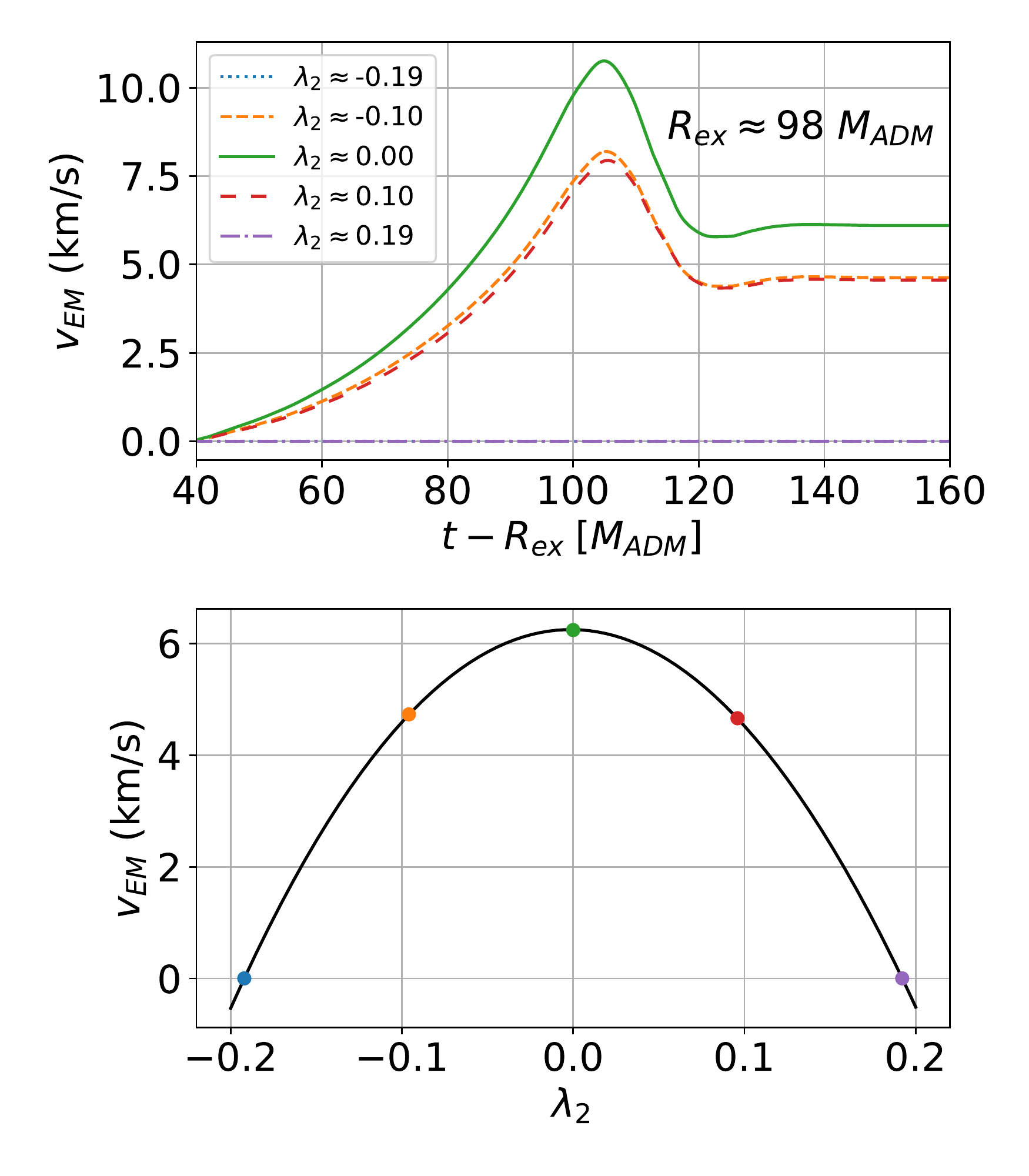}
\caption{Emission of momentum in the EM channel as a function of time (top) and final kick (bottom), for equal-mass binaries ($\rho \approx 1.00$) and $\lambda_1 \approx 0.192$. The solid line is the fitted function of the form~\eqref{eq:fit_form}. The fit error is estimated to be of about 0.6\%.\label{fig:rho1_EM}}
\end{center}
\end{figure}
The actual numerical values for the EM kicks for equal-mass BHs are depicted in Fig.~\ref{fig:rho1_EM}. We define $v = P / M_F$, with $M_F$ being the mass of the final BH as measured by \texttt{QuasiLocalMeasuresEM}. This expression is appropriate in the regime under consideration because the kicks are non relativistic ($v \ll c$). Note that both $P$ and $M_{F}$ have units of mass, so the kick magnitude is independent on the mass-scale of the system. The fitted values from~\eqref{eq:fit_form}  are $A = 169.5 \pm 0.3$ km/s, $B = 0.192 \pm 2 \cdot 10^{-4}$, $C = 0.192 \pm 2 \cdot 10^{-4}$ and $D = 0.03 \pm 0.01$, giving excellent agreement for the zeros of the kick, but a slightly more symmetric curve than the predicted by~\eqref{eqn:thrust}. The uncertainties in the coefficients $B$ and $C$ are of the same order as the variations of $\lambda_1$ between the runs used in the fit.
\begin{figure}[thpb]
\begin{center}
\includegraphics[width=0.43\textwidth]{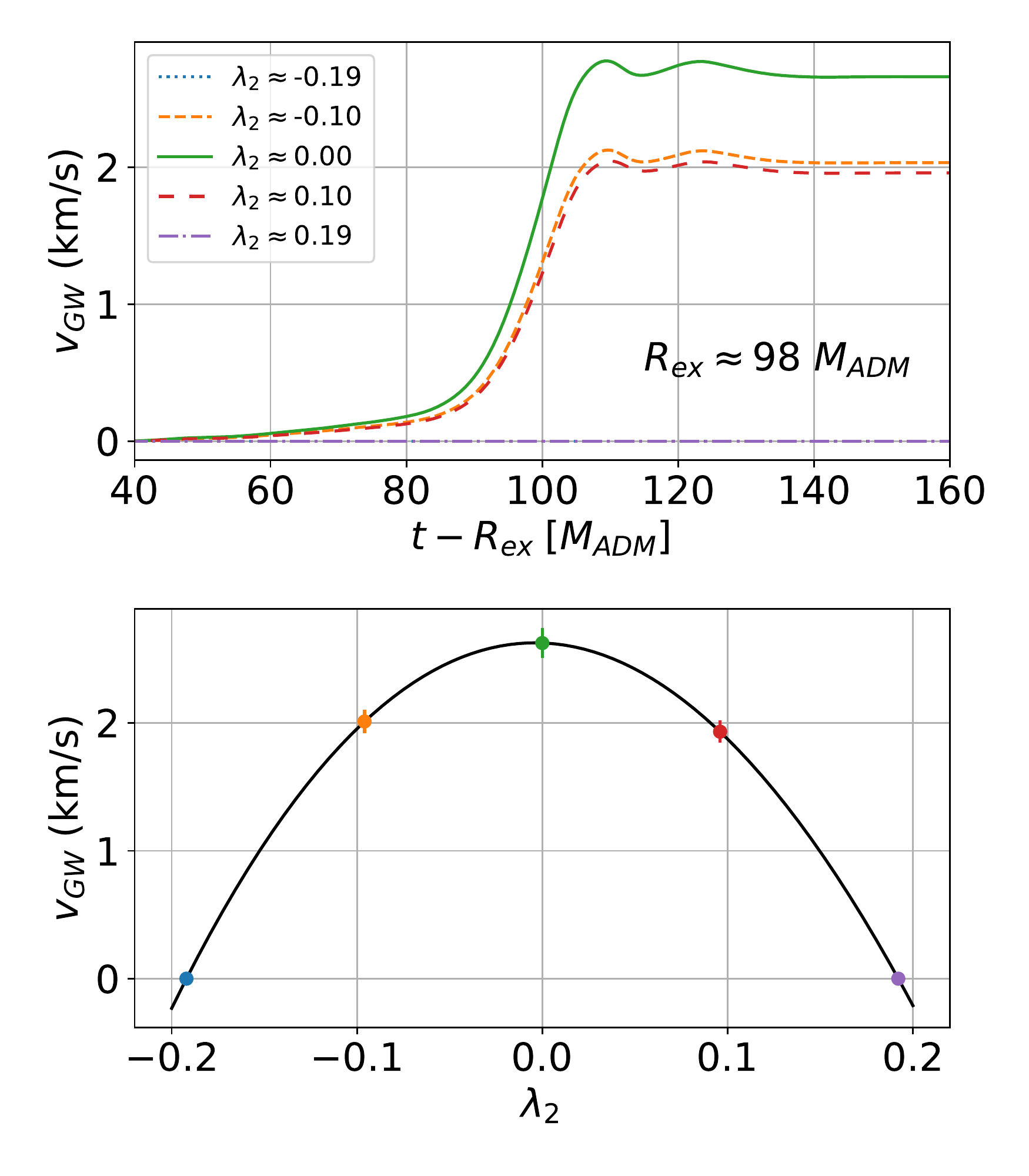}
\caption{Emission of momentum in the GW channel as a function of time (top) and final kick (bottom), for equal-mass binaries ($\rho \approx 1.00$) and $\lambda_1 \approx 0.192$.  The solid line is the fitted function of the form~\eqref{eq:fit_form}. The fit error is estimated to be of about 4.5\%.\label{fig:rho1_GW}}
\end{center}
\end{figure}
Because of the symmetry between the two BHs, we expect the GW kicks for equal mass binaries to vanish for $\lambda_1 = \lambda_2$, which is indeed observed. For other values of $\lambda_2$, however, small GW kicks are produced as a consequence of the EM fields, as shown in Fig.~\ref{fig:rho1_GW}. The profile of the GW kicks as a function of $\lambda_2$ is again almost parabolic, with the fit from~\eqref{eq:fit_form} giving $A = 71.17 \pm 0.03$ km/s, $B = 0.192 \pm 5 \cdot 10^{-5}$, $C = 0.192 \pm 4 \cdot 10^{-5}$ and $D = 0.085 \pm 0.002$.
\begin{figure}[thbp]
\begin{center}
\includegraphics[width=0.43\textwidth]{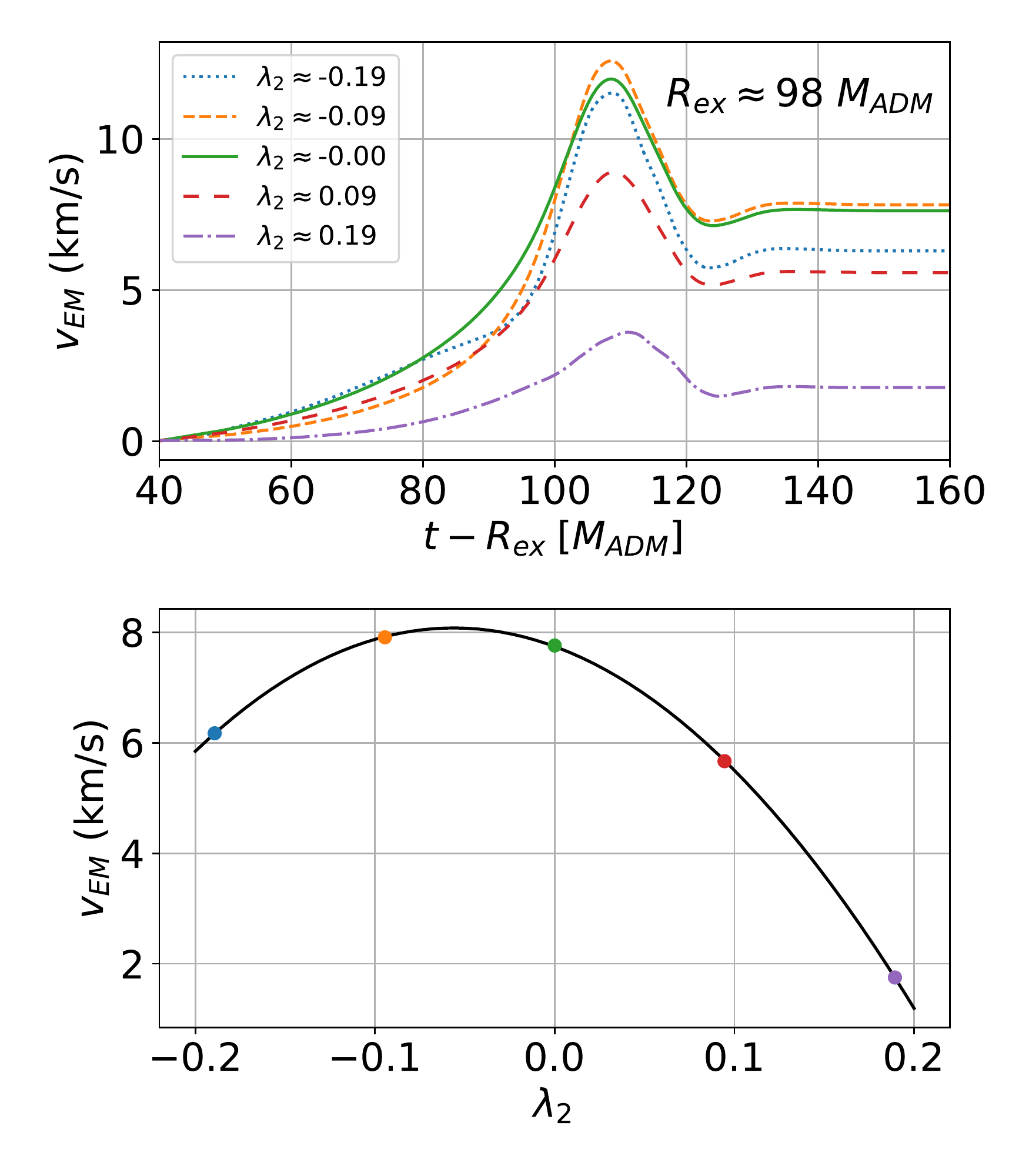}
\caption{EM kicks for mass ratio $\rho \approx 1.94$ and $\lambda_1 \approx 0.195$, showing strong deviation from the prediction of~\eqref{eqn:thrust}, which predicts the zeros to be at $\lambda_2 = \lambda_1, -\lambda_1/\rho$.  The solid line is the fitted function of the form~\eqref{eq:fit_form}. The error is estimated to be of about 0.6\%. \label{fig:rho2_EM}}
\end{center}
\end{figure}

When moving to BH binaries with unequal masses, the numerical results deviate strongly from the Keplerian prediction (see Fig.~\ref{fig:rho2_EM}). In particular, the EM kicks no longer vanish for the zero dipole ($\lambda_2 = \lambda_1$) and zero quadrupole ($\lambda_2 = -\lambda_2/\rho$) configurations, and the maximum is displaced from the predicted position.
\begin{figure}[htbp]
\begin{center}
\includegraphics[width=0.43\textwidth]{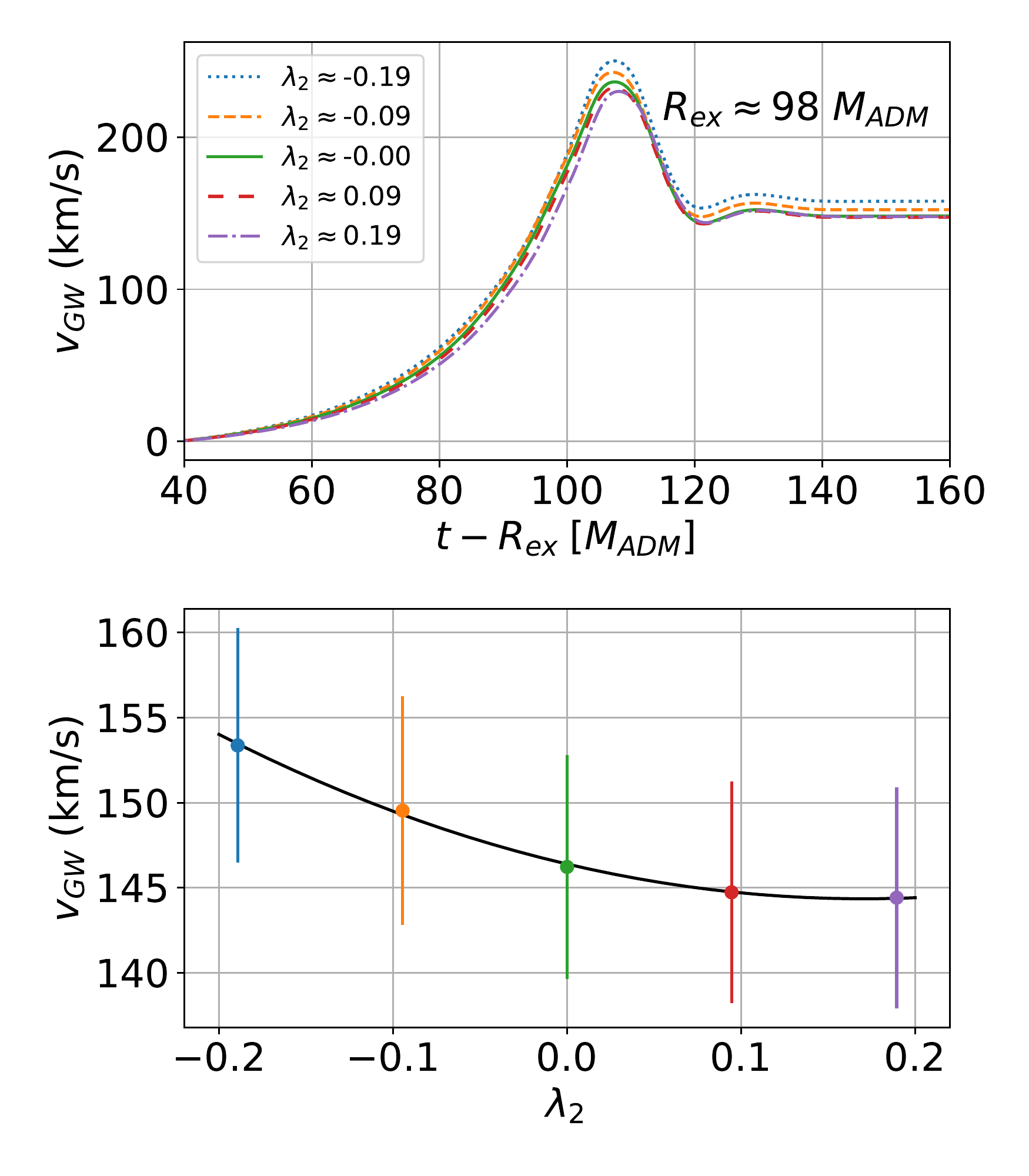}
\caption{Emission of momentum in the GW channel as a function of time (top) and final kick (bottom), for mass ratio $\rho \approx 1.94$ and $\lambda_1 \approx 0.195$.  The solid line is the fitted function of the form~\eqref{eq:fit_form_2}. The error is estimated to be of about 4.5\%.\label{fig:rho2_GW}}
\end{center}
\end{figure}
We do not have, at the present time, a clear explanation of the physical phenomena causing this deviation when $\rho > 1$. We hypothesize, however, that the (much stronger) emission of momentum by the GW channel is ``jiggling'' the system and thus invalidating the assumptions of our derivation and contaminating the EM signal. The GW kicks for mass ratio $\rho \approx 1.94$, plotted in Fig.~\ref{fig:rho2_GW}  can be fitted to be approximately 
\begin{equation}
v_{\rm GW} = v_{\rm GW, \text{neutral}} + A + B{(C - \lambda_2)}^2\,,
\label{eq:fit_form_2}
\end{equation}
with $A = 12.9 \pm 0.2$ km/s, $B = 70 \pm 6$ km/s and $C = 0.17 \pm 0.02$, signaling a quadratic dependence on $\lambda_1 - \lambda_2$. $v_{\rm GW, \text{neutral}} \approx 131.5$ km/s is the value of the kick for a neutral binary with the same mass ratio as the charged ones ($\rho \approx 1.94$). In order to parametrize the effect of the gravitational emission of momentum on the electromagnetic kicks, we fitted an empirical model which adds to the Keplerian profile a contribution proportional to the GW kick as
\begin{equation}
\begin{split}
v_{\rm EM} &= \alpha \frac{ {\left(1 - \lambda_1 \lambda_2\right)}^{5/2}\rho^2 }{{(1+\rho)}^5}  \left(\lambda_1 - \lambda_2\right) \left(\lambda_1 + \rho \lambda_2\right) \\
&+ 2 \beta (1 - \gamma \lambda_2) v_{\rm GW}.
\end{split}
\label{eq:empirical_model}
\end{equation}
\begin{figure}[thpb]
\begin{center}
\includegraphics[width=0.45\textwidth]{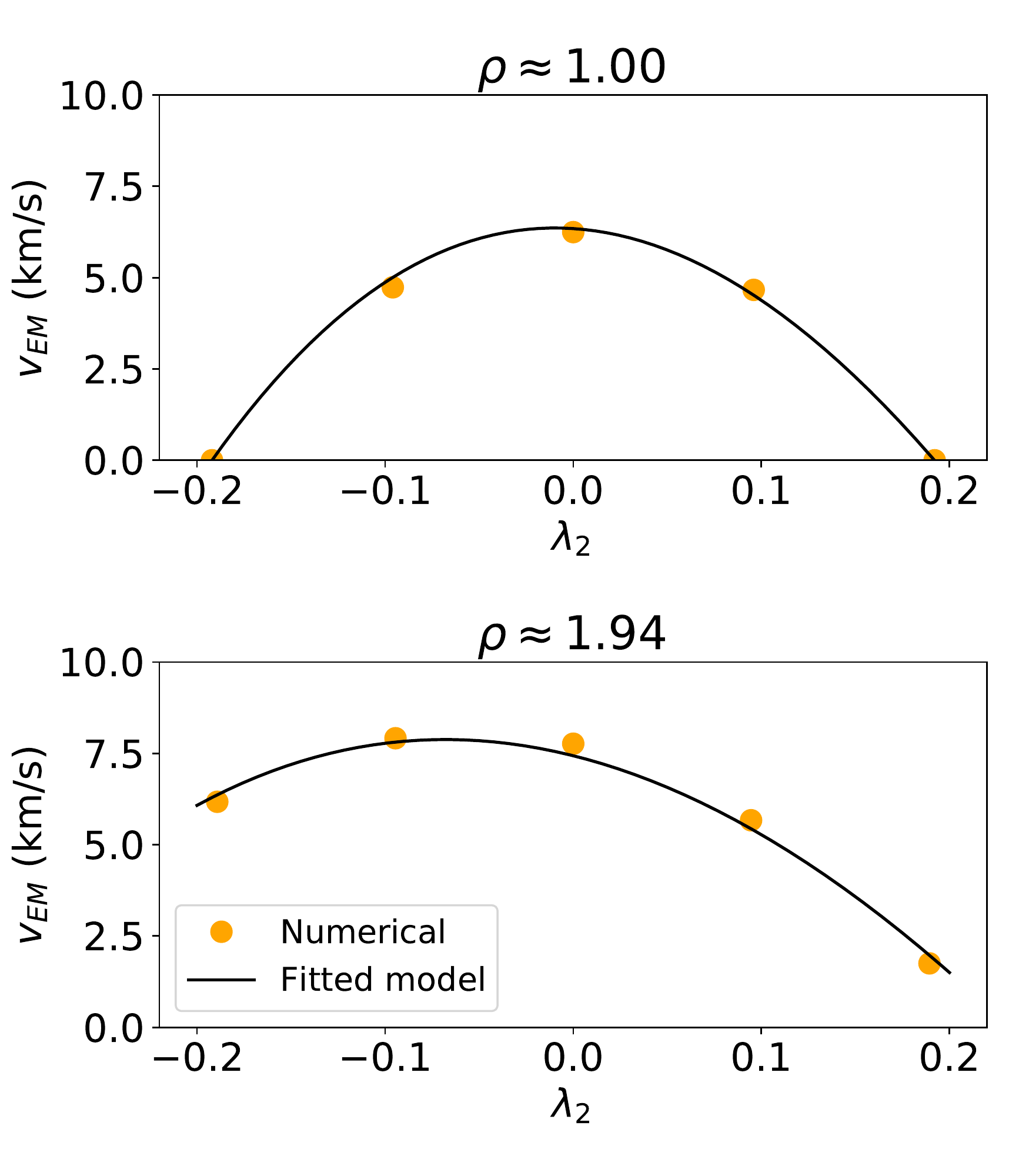}
\caption{Comparison between the EM kicks obtained by numerical simulation and the empirical fitted model~\eqref{eq:empirical_model}. The maximum absolute error is of 0.3 km/s and the maximum relative error (excluding the vanishing kicks) is 13 \%.\label{fig:empirical_fit}}
\end{center}
\end{figure}
By using the results above, we obtain the dimensionless coefficients $\alpha = (5.0 \pm 0.1) \cdot 10^3$ km/s, $\beta = 0.0194 \pm 3 \cdot 10^{-4}$ and $\gamma = 3.5 \pm 0.2$. The comparison of the model with the actual numerical values of the EM kicks is shown in Fig.~\ref{fig:empirical_fit}, giving remarkable accuracy with a maximum error of 0.3 km/s. For the smallest kicks, the relative error can be large (about 13\%). The physical meaning of these parameters is unknown. 
\begin{figure}[thpb]
\begin{center}
\includegraphics[width=0.43\textwidth]{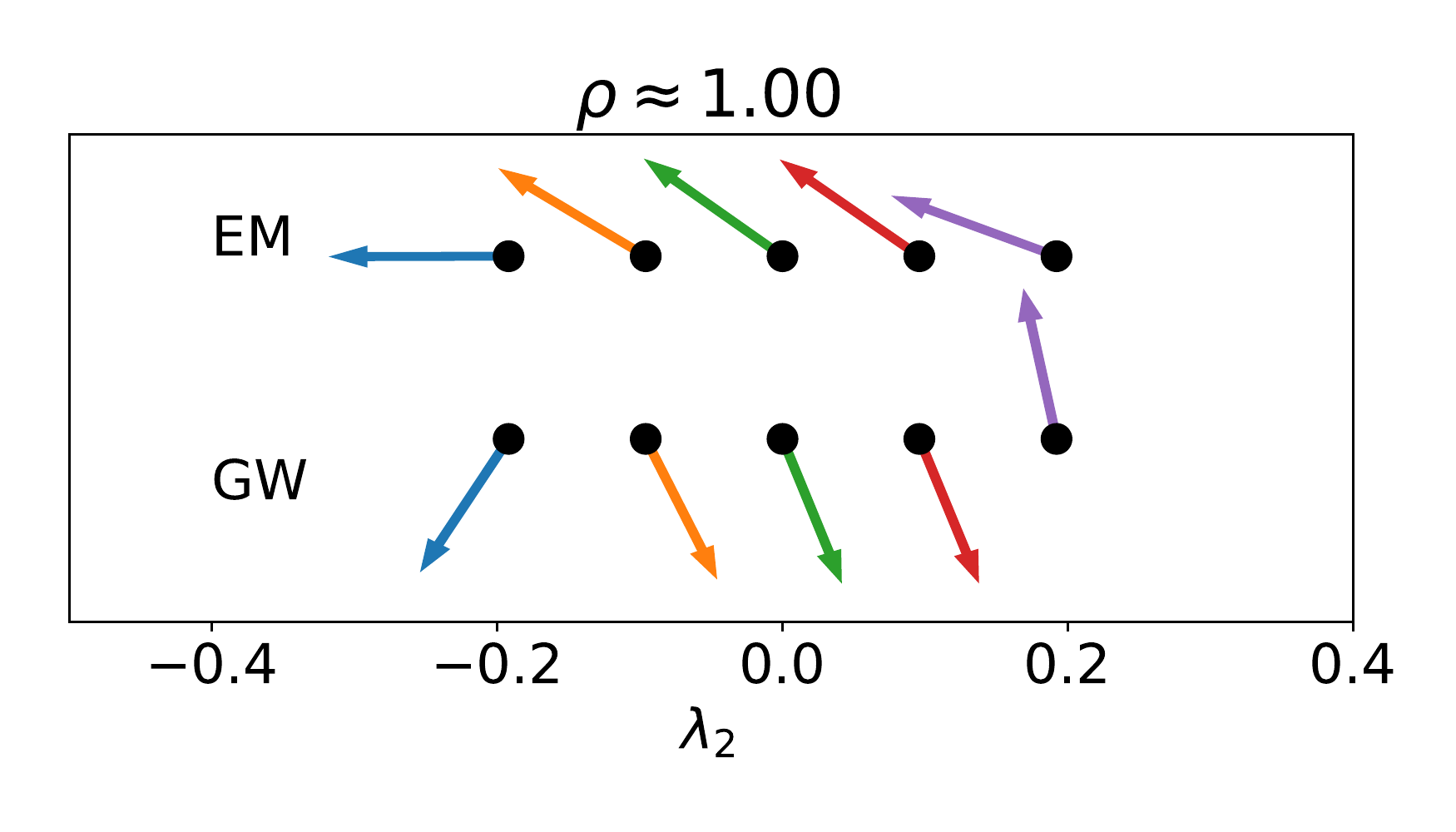} \\
\includegraphics[width=0.43\textwidth]{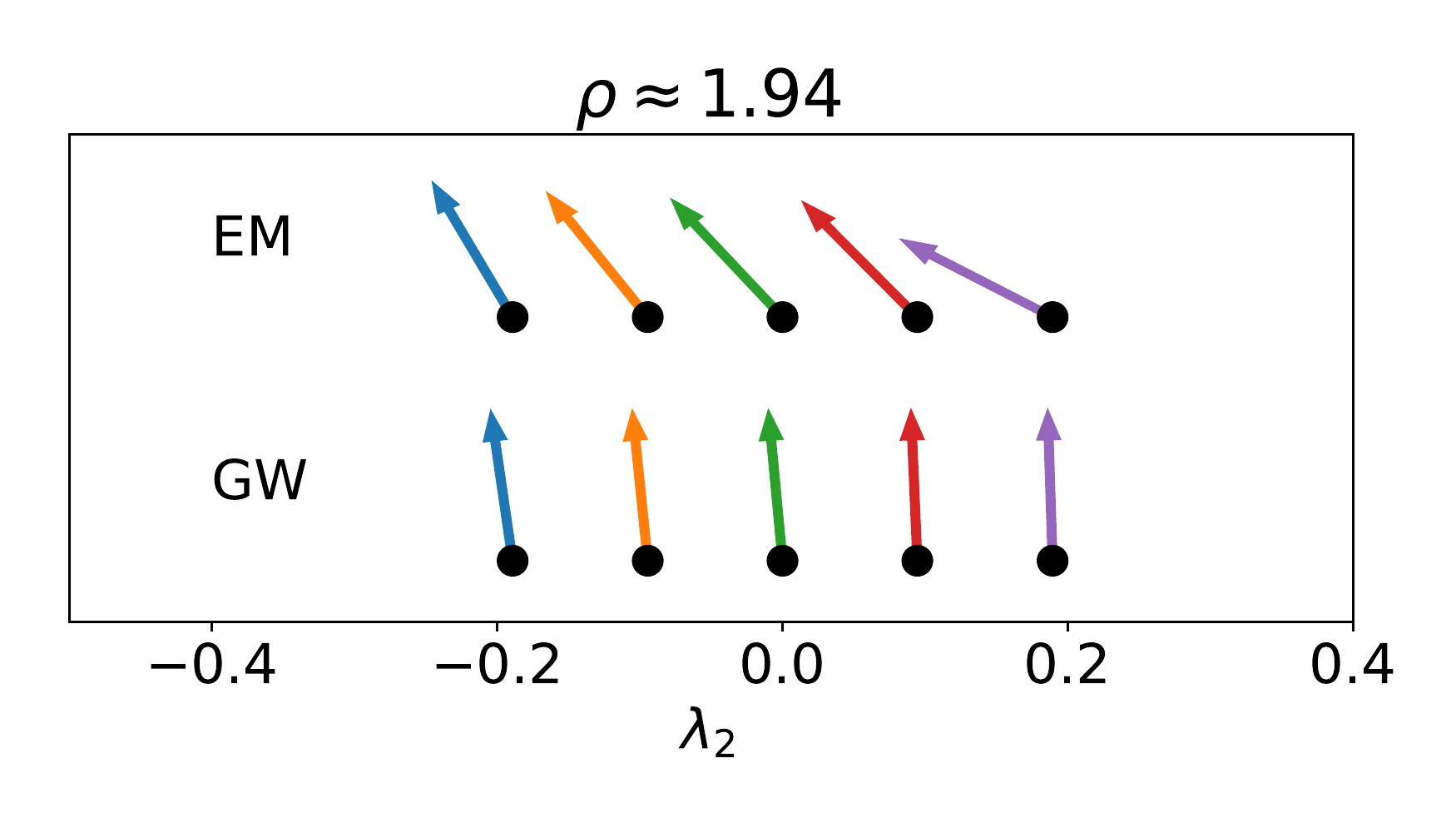}
\caption{Unit vectors showing the directions on the $(x,y)$ orbital plane of the EM and GW kicks for mass ratios 1 and~2. \label{fig:directions}}
\end{center}
\end{figure}
In Fig.~\ref{fig:directions}, the directions of the EM and GW are displayed. It is interesting to notice that for equal masses, the two channels seem to be oriented roughly in opposite directions, while for higher mass ratios the directions seem to be approximately the same.

\section{Discussion}
\label{sec:Discussion}
We computed in a full nonlinear setup the linear momentum carried by EM and GWs in charged BH binaries in quasi-circular orbits, with varying mass ratios and charges. Our results are motivated and also described well by a simple weak-field slow motion expansion, Eq.~\eqref{eq:empirical_model}. Note, however, that we fixed the charge-to-mass ratio of one component, $\lambda_1$. Our fit is not informed by general charge-to-mass ratios, hence it doesn't have a proper neutral limit (notice that when $\lambda_1=\lambda_2=0$, Eq.~\eqref{eq:empirical_model} predicts a nonsensical EM kick).
Our results are robust against changes in the initial orbital separation and small orbital eccentricity. 

We observe that the presence of charge in equal mass binaries can induce a small GW kick at the fully nonlinear level, as long as $|q_1|\neq|q_2|$. For binaries with sufficiently different masses, the EM kick is generally much weaker than its gravitational counterpart for the explored values of the charges. This would make it a subdominant effect in astrophysical scenarios unless the U(1) charge is significantly larger. 

For equal mass binaries, we observe a remarkable agreement with the Keplerian prediction of the general dependence of the EM kicks as a function of the charges. The kick vanishes when either the electric dipole or the electric quadrupole of the system is zero, and presents a maximum close to the middle point between the two roots. The $v(\lambda_2)$ curve is however slightly more symmetrical than expected.

As soon as the masses differ, the $v(\lambda_2)$ curve strongly deviates from the Newtonian prediction, showing nonzero kicks even at zero electric dipole and zero electric quadrupole. We conjecture that this is due to jiggling of the system by the much larger momentum of the gravitational radiation, and we fit a purely phenomenological model that is able to approximately capture the behavior.

The careful exploration of the origin of the deviation is left for future work. The effect of the gravitational momentum emission on the system during the late phases of the inspiral must be further explored, possibly by post-Newtonian approximations, in order to assess the physics behind the numerical results.

\begin{acknowledgments}
We thank the maintainers and developers of the open-source software we used, without which this work would have not been possible.
RL acknowledges financial support provided by FCT/Portugal through the project IF/00729/2015/CP1272/CT0006, and by Next Generation EU though a University of Barcelona Margarita Salas grant from the Spanish Ministry of Universities under the {\it Plan de Recuperaci\'on, Transformaci\'on y Resiliencia}.
GB is supported by NASA Grant 80NSSC20K1542 to the University of Arizona. GB also acknowledges partial support by NSF grant NSF-1759835.
VC is a Villum Investigator and a DNRF Chair, supported by VILLUM FONDEN (grant no.~37766) and by the Danish Research Foundation. VC acknowledges financial support provided under the European
Union's H2020 ERC Advanced Grant ``Black holes: gravitational engines of discovery'' grant agreement
no.\ Gravitas–101052587.
%
This work is supported by NSF grants PHY-1912619 and PHY-2145421 to the
University of Arizona.
MZ acknowledges financial support provided by FCT/Portugal through the IF programme, grant IF/00729/2015, and
by the Center for Research and Development in Mathematics and Applications (CIDMA) through the Portuguese Foundation for Science and Technology (FCT -- Funda\c{c}\~ao para a Ci\^encia e a Tecnologia), references UIDB/04106/2020, UIDP/04106/2020 and the projects PTDC/FIS-AST/3041/2020 and CERN/FIS-PAR/0024/2021.
This project has received funding from the European Union's Horizon 2020 research and innovation programme under the Marie Sklodowska-Curie grant agreement No 101007855.
We thank FCT for financial support through Project~No.~UIDB/00099/2020.
We acknowledge financial support provided by FCT/Portugal through grants PTDC/MAT-APL/30043/2017 and PTDC/FIS-AST/7002/2020.
This work has also received funding from MICINN grant PID2019-105614GB-C22, AGAUR grant 2017-SGR 754, and State Research Agency of MICINN through the ``Unit of Excellence Mar\'ia de Maeztu 2020-2023'' award to the Institute of Cosmos Sciences (CEX2019-000918-M). Work supported by the Spanish Agencia Estatal de Investigaci\'on (Grants PGC2018-095984-B-I00 and PID2021-125485NB-C21).
We acknowledge that the results of this research have been achieved using the DECI resource Snellius based in The Netherlands at SURF with support from the PRACE aisbl,
the Navigator cluster, operated by LCA-UCoimbra, through project~2021.09676.CPCA, as well as the ``Baltasar-Sete-Sóis'' cluster at IST.
Results were cross-checked with \texttt{kuibit}~\cite{Bozzola:2021hus}, which is
based on \texttt{NumPy}~\cite{NumPy}, \texttt{SciPy}~\cite{SciPy}, and
\texttt{h5py}~\cite{h5py}. GB and VP thank Maggie Smith for cross-checking Equation~\eqref{eq:Pem}.

\end{acknowledgments}

\appendix

\section{Multipolar Expansion}%
\label{app:Multipoles}

We extend to arbitrary spin weights the approach used in~\cite{Ruiz:2007yx} to express the energy and momentum emission in terms multipole moments of the Newman-Penrose scalars $\Psi_{4}$ and $\Phi_{2}$. Let $_{s}Y^{l,m}(\theta, \varphi)$ be the spin-weighted spherical harmonic with spin $s$ and angular numbers $l$ and $m$, we have that
\begin{equation}
\begin{split}
\Psi_4 &= \sum_{l=2}^\infty \sum_{m=-l}^l A^{l,m}(t) \; _{-2}Y^{l,m}(\theta, \varphi), \\
\Phi_2 &= \sum_{l=1}^\infty \sum_{m=-l}^l B^{l,m}(t) \; _{-1}Y^{l,m}(\theta, \varphi)\,.
\end{split}
\end{equation}

\subsection{Energy Emission}

In this case, the solution is immediate thanks to the orthogonality relation between spin-weighted spherical harmonics
\begin{equation}
\oint {}_{s}Y^{l,m} {}_{s'}\bar Y^{l',m'} d\Omega = \delta_{ss'}\delta_{ll'}\delta_{mm'},
\end{equation}
where the bar over complex quantities denotes complex conjugation. Therefore, equations (\ref{eq:Egw}) and (\ref{eq:Eem}) can be simply rewritten as
\begin{equation}
\begin{split}
\frac{dE_{\rm GW}}{dt} & = \lim_{r \to \infty} \frac{r^2}{16 \pi} \sum_{l,m}\left| \int_{-\infty}^t A^{l,m}dt'\right|^2, \\
\frac{dE_{\rm EM}}{dt} & = \lim_{r \to \infty} \frac{r^2}{4 \pi} \sum_{l,m}\left|B^{l,m}\right|^2.
\end{split}
\end{equation}
The time integrations are computed numerically using the composite trapezoidal rule. See Section \ref{sec:evolutions} for more details on the treatment of the initial ``junk'' radiation.

\subsection{Momentum Emission}

In this case, the derivation of the expressions is slightly more involved due to the presence of the radially outward-pointing unit vector $n^i$ in the integrals which does not allow for direct application of the orthogonality relation. It can be simplified, however, by noticing that $n^i$ can be written in terms of the ordinary spherical harmonics (spin weight 0) as
\begin{equation}
\begin{split}
n^x &= \sin\theta \cos \varphi = \sqrt{\frac{2\pi}{3}} \left[_0Y^{1, -1} - _0Y^{1, 1}\right],\\
n^y &= \sin\theta \sin \varphi = i\sqrt{\frac{2\pi}{3}} \left[_0Y^{1, -1} + _0Y^{1, 1}\right],\\
n^z &= \cos\theta = 2\sqrt{\frac{\pi}{3}} Y^{1,0},
\end{split}
\end{equation}

and the calculation becomes even simpler in terms of the complex quantity 
\begin{equation}
n^+ \equiv n^x + i n^y = -\sqrt{\frac{8\pi}{3}} \; _0Y^{1, 1}.
\end{equation}
Using the expression of the angular integral of a triple product of spin-weighted spherical harmonics in terms of the Wigner 3-$lm$ symbols, and after some algebra, we can define the spin-weighted coefficients 

\begin{align*}
{}_{s}a_{l,m} & = -s\frac{\sqrt{(l - m)(l + m + 1)}}{2l(l+1)}, \\
{}_{s}b_{l,m} & = \frac{1}{2l}\sqrt{\frac{(l+s)(l-s)(l + m)(l + m - 1)}{(2l-1)(2l+1)}}, \\
{}_{s}c_{l,m} & = -\frac{s\;m}{l(l+1)}, \\
{}_{s}d_{l,m} & = \frac{1}{l}\sqrt{\frac{(l+s)(l-s)(l-m)(l+m)}{(2l-1)(2l+1)}},
\end{align*}
which lead to the final expressions for momentum emission
\begin{widetext}
\begin{equation}
\begin{split}
\frac{dP^{+}_{\rm GW}}{dt} & = \lim_{r \to \infty} \frac{r^2}{8 \pi} \sum_{l,m}\left[ \int_{-\infty}^t A^{l,m}dt'\right] \left[\int_{-\infty}^t  \left({}_{-2}a_{l,m} \bar A^{l,m+1} + {}_{-2}b_{l,-m} \bar A^{l-1,m+1} - {}_{-2}b_{l+1,m+1} \bar A^{l+1,m+1} \right) dt' \right], \\
\frac{dP_{\rm GW}^{z}}{dt} & = \lim_{r \to \infty} \frac{r^2}{16 \pi} \sum_{l,m} \left[ \int_{-\infty}^t A^{l,m}dt'\right] \left[ \int_{-\infty}^t  \left( {}_{-2}c_{l,m} \bar A^{l,m} + {}_{-2}d_{l,m} \bar A^{l-1,m} + {}_{-2}d_{l+1,m} \bar A^{l+1,m} \right) dt'\right], \\
\frac{dP_{\rm EM}^+}{dt} & = \lim_{r \to \infty} \frac{r^2}{2 \pi} \sum_{l,m} B^{l,m} \left[ {}_{-1}a_{l,m} \bar B^{l,m+1} + {}_{-1}b_{l,-m} \bar B^{l-1,m+1} - {}_{-1}b_{l+1,m+1} \bar B^{l+1,m+1} \right], \\
\frac{dP_{\rm EM}^z}{dt} & = \lim_{r \to \infty} \frac{r^2}{4 \pi} \sum_{l,m} B^{l,m} \left[ {}_{-1}c_{l,m} \bar B^{l,m} + {}_{-1}d_{l,m} \bar B^{l-1,m} + {}_{-1}d_{l+1,m} \bar B^{l+1,m} \right].
\end{split}
\end{equation}
\end{widetext}

\section{Numerical Checks}%
\label{app:Checks}

\subsection{Neutral Kicks}

\begin{figure}[thbp]
\begin{center}
\includegraphics[width=0.45\textwidth]{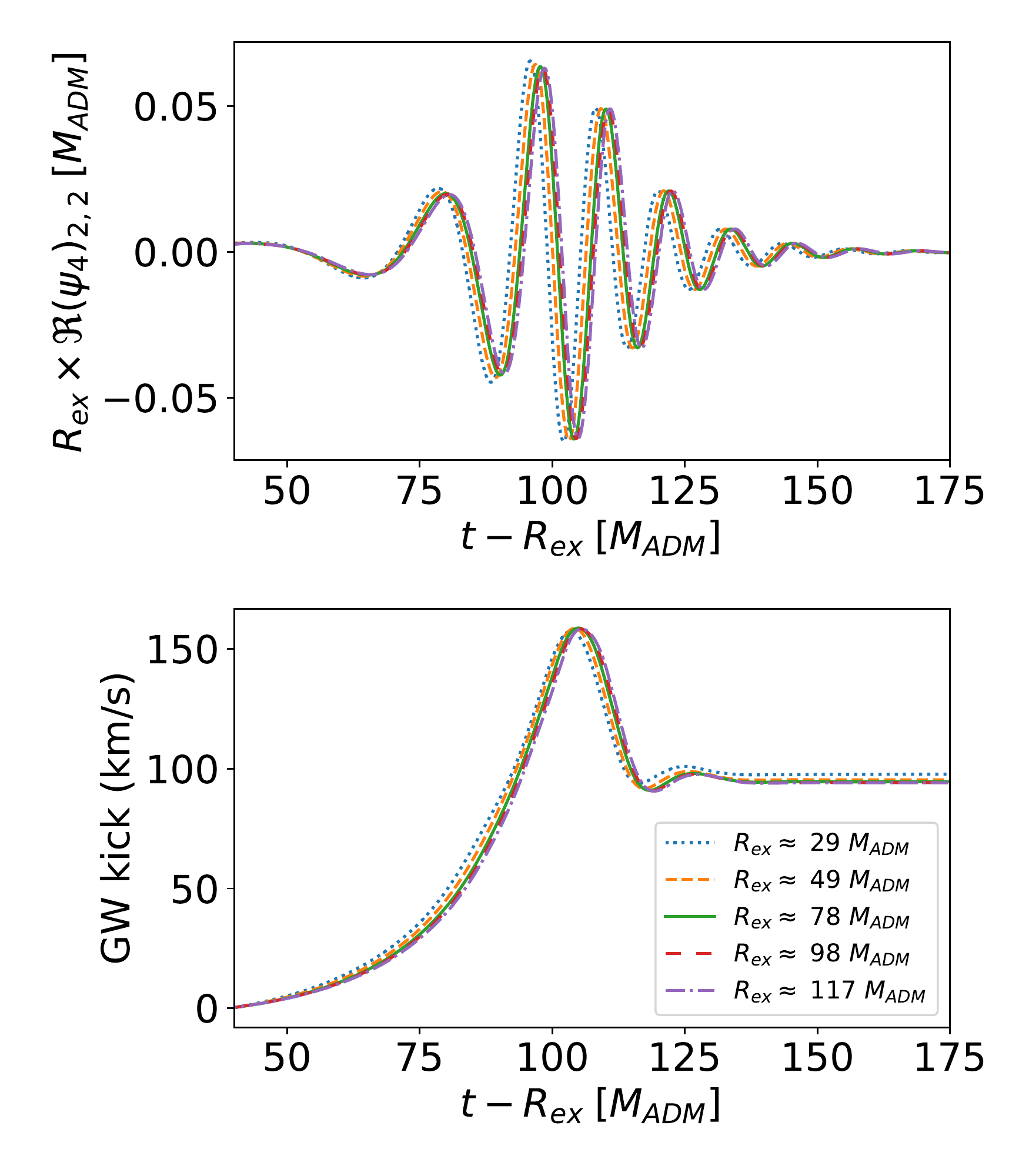}
\caption{Multipolar $l = 2$, $m = 2$ mode of $\Re(\Psi_4)$ and velocity profiles as a function of extraction radius for a neutral binary with mass ratio 1.43.\label{fig:neutral_waveform}}
\end{center}
\end{figure}

\begin{figure}[htbp]
\begin{center}
\includegraphics[width=0.45\textwidth]{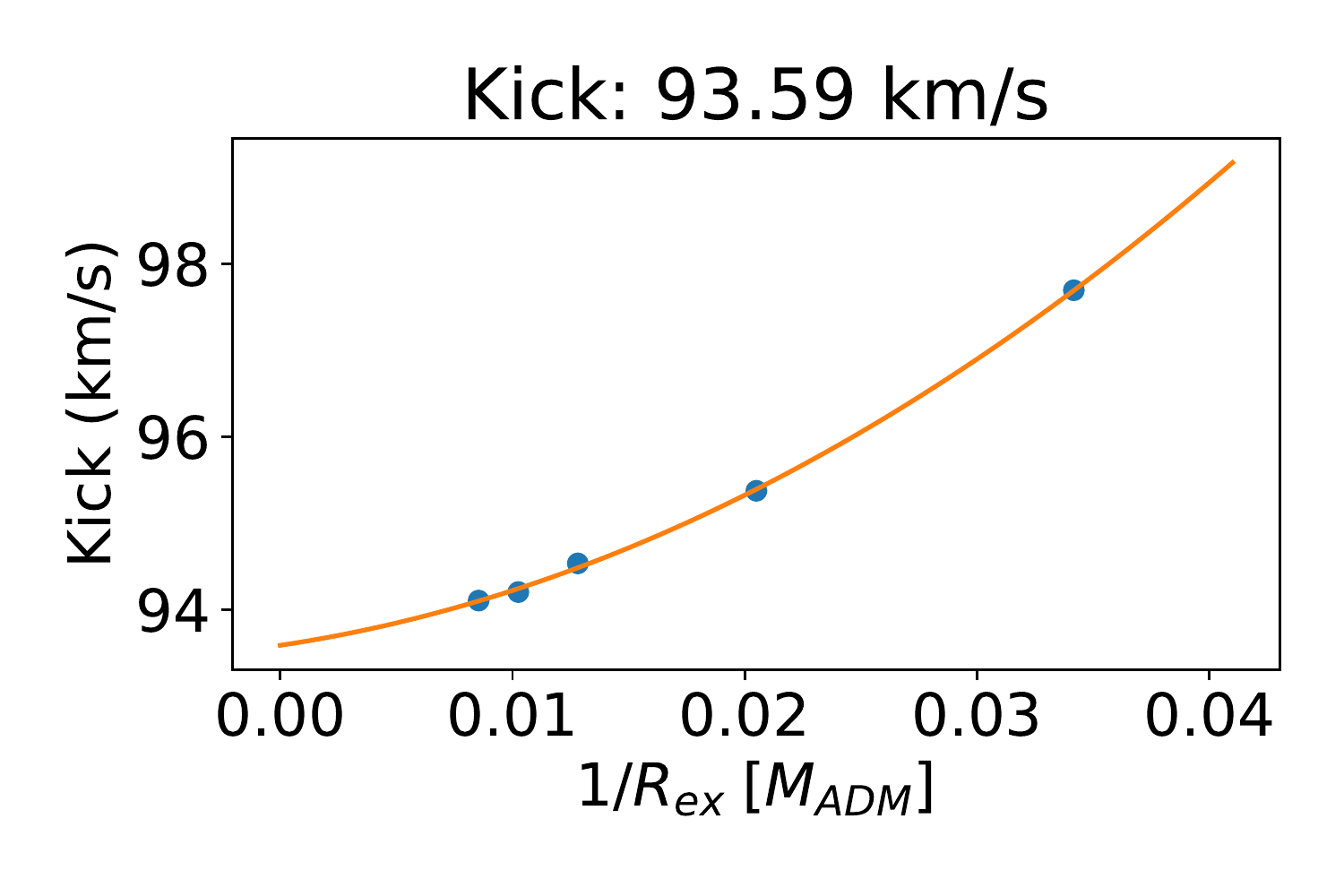}
\caption{Extrapolation to $R_{\rm ex} \to \infty$ to estimate the kick, via a second order polynomial, for a neutral binary with mass ratio 1.43.\label{fig:kick_extrapolation}}
\end{center}
\end{figure}

We performed consistency checks  the numerical results to quantify the quality of the simulations. First, we reproduced established results in the context of kicks in neutral BH. In~\cite{Gonzalez:2006md}, simulation data was used to estimate the phenomenological trend for momentum lost by GWs
\begin{equation}
\begin{split}
v_{\rm GW} &\approx 1.2 \times 10^4 \eta^2 \sqrt{1-4\eta} (1-0.93\eta), \\ \eta &= \rho/{(1+\rho)}^{2}.
\label{eq:neutral_fit}
\end{split}
\end{equation}
Our simulations reproduce this result. Figure~\ref{fig:neutral_waveform} shows the real part of the $(l,m) = (2,2)$ component of $\Psi_4$ (multiplied by the extraction radius $R_{\rm ex}$) and the GW kick of a neutral binary with $\rho = 1.43$ (as measured by \texttt{QuasiLocalMeasuresEM}). For this mass ratio,~\eqref{eq:neutral_fit} predicts a kick of 95.92 km/s. In Fig.~\ref{fig:kick_extrapolation} we obtain a value of 93.59 km/s. This is an error of 2.5\% with respect to~\eqref{eq:neutral_fit}, within the error bar given in~\cite{Gonzalez:2006md}.

\subsection{Convergence of the Multipolar Expansion}

\begin{figure}[htbp]
\begin{center}
\includegraphics[width=0.45\textwidth]{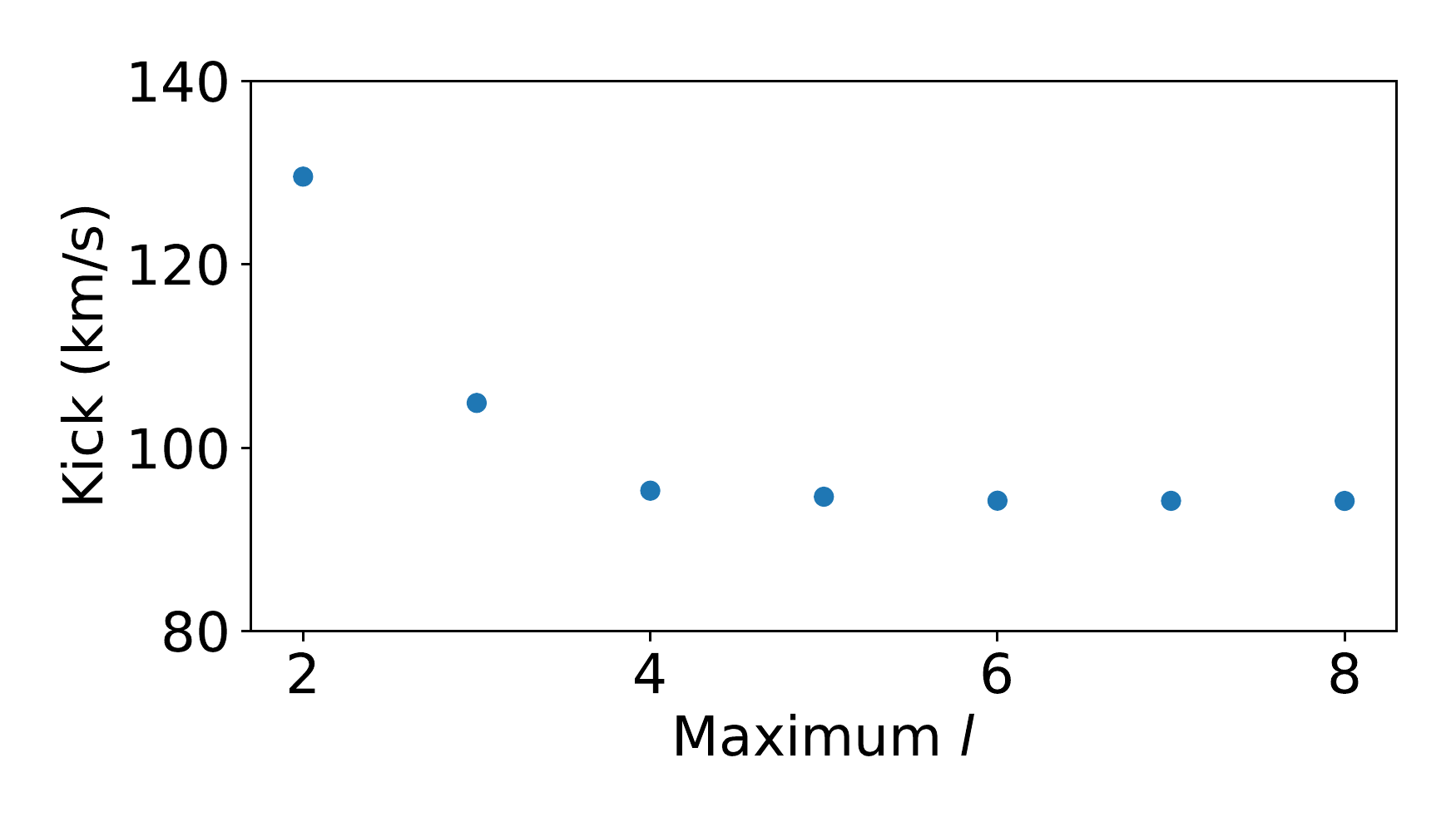}
\caption{Value of the GW kick computed to different orders in the multipole expansion, for a neutral binary with mass ratio 1.43. The extraction radius is $R_{\rm ex} \approx 98 M_{ADM}$.\label{fig:l_convergence}}
\end{center}
\end{figure}
The kicks have been extracted from the multipolar expansion coefficients of the NP scalars, up to $l = 8$, as explained in Appendix~\ref{app:Multipoles}. In Fig.~\ref{fig:l_convergence} we show the value of the extracted kick computed using different numbers of multipole coefficients. Although the contribution to the total kick is dominated by the $(l,m) = (2,2)$ mode, the value experiences significant changes until $l = 4$. It is possible to see from the plots that the waveform and the velocity profiles converge as $R_{\rm ex}$ get larger. The same happens with the EM sector for charged binaries. More quantitatively, we can plot the final kick as a function of $1/R_{\rm ex}$ and extrapolate to obtain the value as $R_{\rm ex} \to \infty$.

\subsection{Self-Convergence Testing}

\begin{figure}[htbp]
\begin{center}
\includegraphics[width=0.45\textwidth]{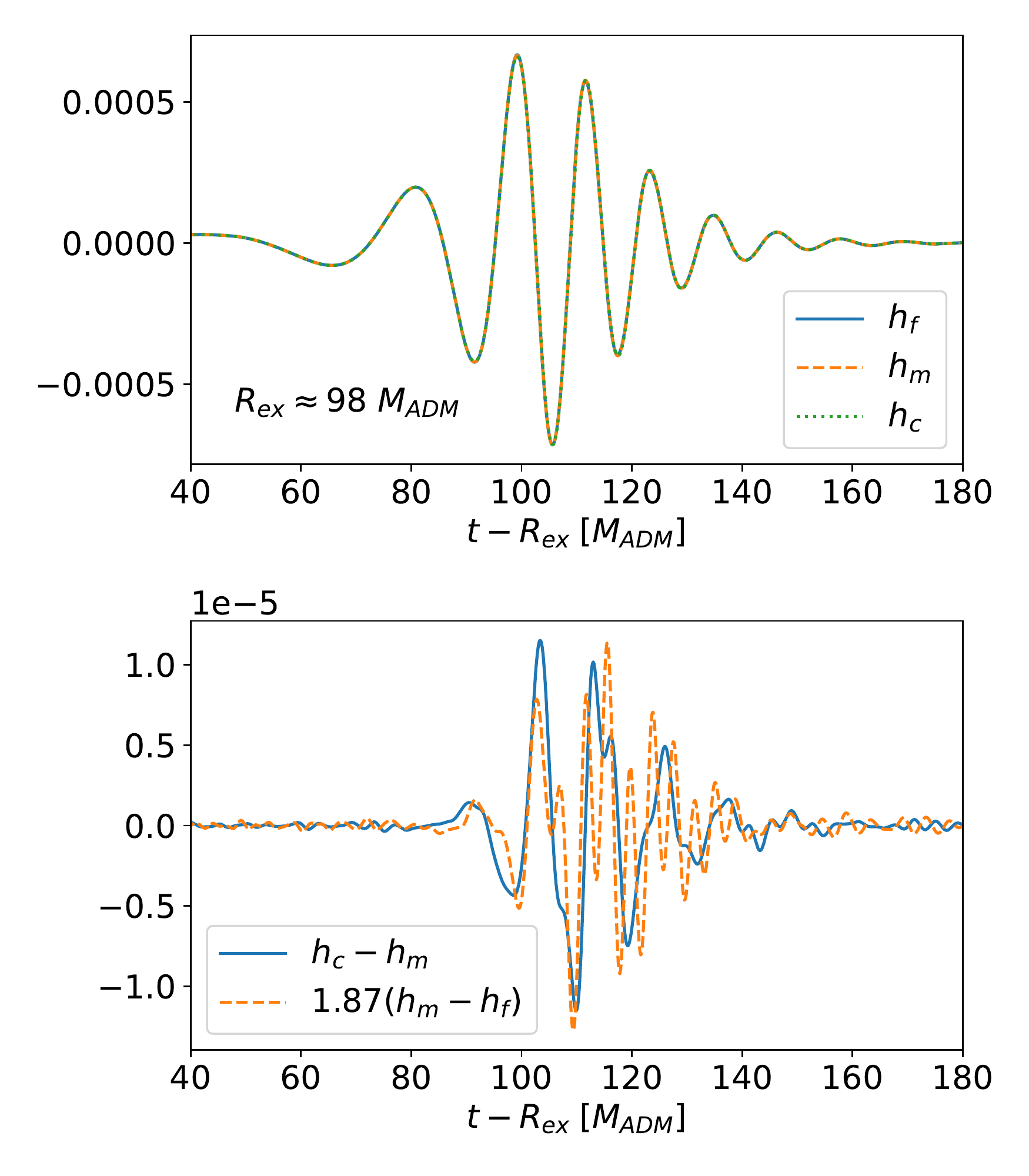}
\caption{Self-convergence testing on the $(l,m) = (2,2)$ mode of Newman-Penrose scalar $\psi_4$; results are compatible with 4th order convergence.\label{fig:convergence}}
\end{center}
\end{figure}
Figure~\ref{fig:convergence} shows a self-convergence analysis performed on equal-mass binary BHs with $\lambda_1 = 0.2, \; \lambda_2 = 0$, by using three different resolutions (fine, medium, and coarse). Each run has 8 refinement levels (each one halving the previous grid spacing), with outermost grid spacings of $h_c = 1.47$, $h_m = 1.23$ and $h_f = 0.99$ in units of $M_{ADM}$. This produces the scaling factor of 1.87 for 4th order convergence, computed from $(h_c^4 - h_m^4)/(h_m^4 - h_f^4) \approx 1.87$. We observe high frequency noise as described in~\cite{Bozzola:2020mjx}.

\subsection{Initial Conditions}

Starting the simulations with the orbiting BHs at a finite (and small) distance will always introduce some systematic error to the computation of the kicks, since we are neglecting all emission from the system prior to the computational starting time. Additionally, the initial momenta of a quasi-circular inspiral have to be estimated, and will also contain deviations from the momenta that would be achieved at that distance when inspiraling from an infinite distance.
For this reason, we perform a check by repeating one of the inspiral cases in this paper (namely $\rho \approx 1.94$, $\lambda_1 = 0.195$ and $\lambda_2 = -0.09$), but starting at a distance of $d = 7.84 M_{ADM}$ instead of $d = 6.86 M_{ADM}$. This results in a difference of 0.25\% in the EM kick and 4\% in the GW kick. This is of the same order as the error found in the previous section. We therefore estimate that our kicks have errors of order few percent.

\bibliography{references}

\end{document}